\documentstyle[12pt,epsfig]{article}
 1
\def\as{\alpha_{\rm S}}

\def\citenum#1{{\def\@cite##1##2{##1}\cite{#1}}}
\def\citea#1{\@cite{#1}{}}

\def\as{\alpha_{\rm S}}

\def\l{\lambda}

\def\s{\sigma}

\def\({\left(}
\def\){\right)}

\def\citenum#1{{\def\@cite##1##2{##1}\cite{#1}}}
\def\citea#1{\@cite{#1}{}}

\def\l1vt{\vec{l_{1\perp}}}

\def\bt{b_{\perp}}

\def\bt2{$b^2_t$}

\def\jol1{$J_0(\,l_{1\perp}\,r_{\perp}\,)$}

\def\citea#1{\@cite{#1}{}}

\def\beq{\begin{equation}}
\def\eeq{\end{equation}}
\def\bea{\begin{eqnarray}}
\def\eea{\end{eqnarray}}

\def\eq#1{{Eq.~(\ref{#1})}}

\def\bbbz{{\mathchoice {\hbox{$\sf\textstyle Z\kern-0.4em Z$}}
{\hbox{$\sf\textstyle Z\kern-0.4em Z$}}
{\hbox{$\sf\scriptstyle Z\kern-0.3em Z$}}
{\hbox{$\sf\scriptscriptstyle Z\kern-0.2em Z$}}}}
\def\npb#1#2#3{    {\it Nucl. Phys. }{\bf B#1} (19#2) #3}
\def\plb#1#2#3{    {\it Phys. Lett. }{\bf B#1} (19#2) #3}
\def\prd#1#2#3{    {\it Phys. Rev. }{\bf D#1} (19#2) #3}

\def\zpc#1#2#3{    {\it Z. Phys. }{\bf C#1} (19#2) #3}

\def\sjnp#1#2#3{   {\it Sov. J. Nucl. Phys. }{\bf #1} (19#2) #3}

\relax
\begin{document}
\newcounter{savefig}
\newcommand{\alphfig}{\addtocounter{figure}{1}%
\setcounter{savefig}{\value{figure}}%
\setcounter{figure}{0}%
\renewcommand{\thefigure}{\mbox{\arabic{savefig}-\alph{figure}}}}
\newcommand{\resetfig}{\setcounter{figure}{\value{savefig}}%
\renewcommand{\thefigure}{\arabic{figure}}}

\begin{titlepage}
\noindent
\begin{flushright}
December   1997 \\ 
TAUP 2471/97
\end{flushright}
\vspace{1cm}
\begin{center}
{\Large \bf  THE  $\mathbf{ F_2}$  SLOPE 
  AND
SHADOWING}\\[2ex]
{\Large \bf  CORRECTIONS  IN DIS}\\[4ex]

{\large E. G O T S M A N${}^{a), 1)}$, E. L E V I N${}^{a),b),2)}$\,\,
 and U.\,\,M A O R${}^{a), 3)}$}
 \footnotetext{$^{1)}$ Email: gotsman@post.tau.ac.il .}
\footnotetext{$^{2)}$ Email: leving@ccsg.tau.ac.il.}
\footnotetext{$^{3)}$ Email: maor@ccsg.tau.ac.il.}\\[4.5ex]
{\it a) School of Physics and Astronomy}\\
{\it Raymond and Beverly Sackler Faculty of Exact Science}\\
{\it Tel Aviv University, Tel Aviv, 69978, ISRAEL}\\[1.5ex]
{\it b)  Theory Department, Petersburg Nuclear Physics Institute}\\
{\it 188350, Gatchina, St. Petersburg, RUSSIA}\\[3.5ex]
\end{center}
~\,\,\,
\vspace{2cm}

{\large \bf Abstract:}

Recent  HERA  low $Q^2$ data show that the logarithmic  slope of the
proton
structure function (  $\frac{\partial F_2}{\partial \ln Q^2}$ )    
 is significantly different  from perturbative QCD expectations   for
small
values of
 $Q^2$ at exeedingly small values of $x$. We show that shadowing (
screening ) corrections  provide a natural  explanation for this
experimental observation.

\end{titlepage}

 Recent  HERA data  on $Q^2$  and  $x$  dependence of the
logarithimic   $Q^2$  derivative of the proton  structure function
$F_2(x,Q^2)$,  $\frac{\partial F_2}{\partial \ln Q^2}$
,  have become available \cite{HERALOWQ}\cite{F2SLHERA}. This is shown in
Fig.1 where
each point corresponds to a different  values of  $Q^2$ and $x$. As
seen,
 this observable rises steeply  with decreasing $x$ to values of $x$ of
approximately $10^{-4}$  with values of  $Q^2 $ larger than a few
$ GeV^2$.
 However,    in
the exeeedingly
small  $x \,<\, 10^{-4}$ and low $Q^2\,<\,5 \,GeV^2$ values, the
behaviour  of the $F_2$ slope is completely different - decreasing rapidly
with   $x $ and/or $Q^2$  getting smaller( see Fig.1). This
dramatic
change of
$\frac{\partial F_2}{\partial \ln Q^2}$ is not compatible with the
prediction  of perturbative QCD (pQCD ). This is shown  in Fig.1 where the
 HERA  data \cite{HERALOWQ} is compared with the theoretical DGLAP
expectations  \cite{DGLAP} based on the GRV'94 parton distribution input
\cite{GRV}. Specifically, in the small $x$ limit of the DGLAP equations we
have 
\beq \label{DGLAPSL}
\frac{\partial F^{DGLAP}_2(x,Q^2)}{\partial
\ln(Q^2/\Lambda^2)}\,\,=\,\,\frac{2 \as}{9 \pi} xG^{DGLAP}(x,Q^2)\,\,,
\eeq
where the expected rise of the $F_2$ slope is associated  with the
logarithmic $Q^2$ growth of $xG(x,Q^2)$ implied by the DGLAP equations in
the small $x$ limit.

The new data pose a severe theoretical 
 challenge as it requires a successful description of DIS in 
 the transition region
characterized by intermediate distances. These distances are smaller
than the confinement scale $\frac{1}{\Lambda}$, where
$\as(\Lambda^2)\,\gg\,1$, but are still  not small enough to justify a
reliable pQCD  calculation.
For sufficiently small $Q^2$ we expect non-perturbative contributions to
be important in DIS. However, if the transition region corresponds to
larger  values of $Q^2$ as it appears  from Fig.1, we presume
that
these nonperturbative contributions to DIS will  depend on several average
characteristics such as 
vacuum expectation of the square of gluon tensor and  the
correlatrion length
of two gluons within  the hadron. These properties are reflected in  the
QCD
Sum Rules
approach \cite{SRQCD}.

In order to assess the significance of the new data we examine both
$\frac{\partial F_2}{\partial \ln Q^2}$  and  $F_2$. 
A change in the functional $Q^2$ dependence of  $\frac{\partial
F_2}{\partial \ln Q^2}$  at small $Q^2$ values can be deduced  from
general arguments.
 Since  $\s_t\,=\,\frac{4 \pi^2
\alpha^{em}}{Q^2} F_2$ is finite ( non zero )\, for real
photoproduction\,\, (
$Q^2$ = 0 and $x$ =0 ), we conclude that at small $Q^2$ values 
 $F_2 \,\propto\,Q^2$   (  see Ref. \cite{ALLM} for  a more detail discussion).
Accordingly, an eventual change in the functional $x$ dependence of 
$\frac{\partial F_2}{\partial \ln Q^2}$ in the limit of exceedingly small
$x$ is expected. However, the particular   $x$ and $Q^2$ location of this
change,  its dynamical mechanism and the relevance to pQCD are not
{ \it apriori} clear. We note  that the $F_2$ data in the HERA
kinematic region at low $x$ and  $Q^2 \,\geq\, 1\,GeV^2$ are well
reproduced by the DGLAP
evolution equations with input parton distributions \cite{GRV} \cite{MRS}
\cite{CTEQ} starting from very low $Q^2$ values. The importance of the new 
$\frac{\partial F_2}{\partial \ln Q^2}$ data is that it opens a new window
through which a transition region is observed in the  $Q^2$ range of $1
- 4 \,GeV^2$ and $ x \,<\,10^{-4}$. This transition is apparently not
resolved through the study of $F_2$ in the same $x$ and $Q^2$ domain. This
transition is not  predicted by the DGLAP evolution ( see \eq{DGLAPSL} )
as can readily be seen in Fig.1.

The objective of this letter is to show that shadowing ( screening
)
corrections 
\newline
( SC ) to $F_2$ account for the
deviation
from the DGLAP
predictions and provide a natural explanation for the observed
experimental phenomena.

\begin{figure}[htbp]
\epsfig{file=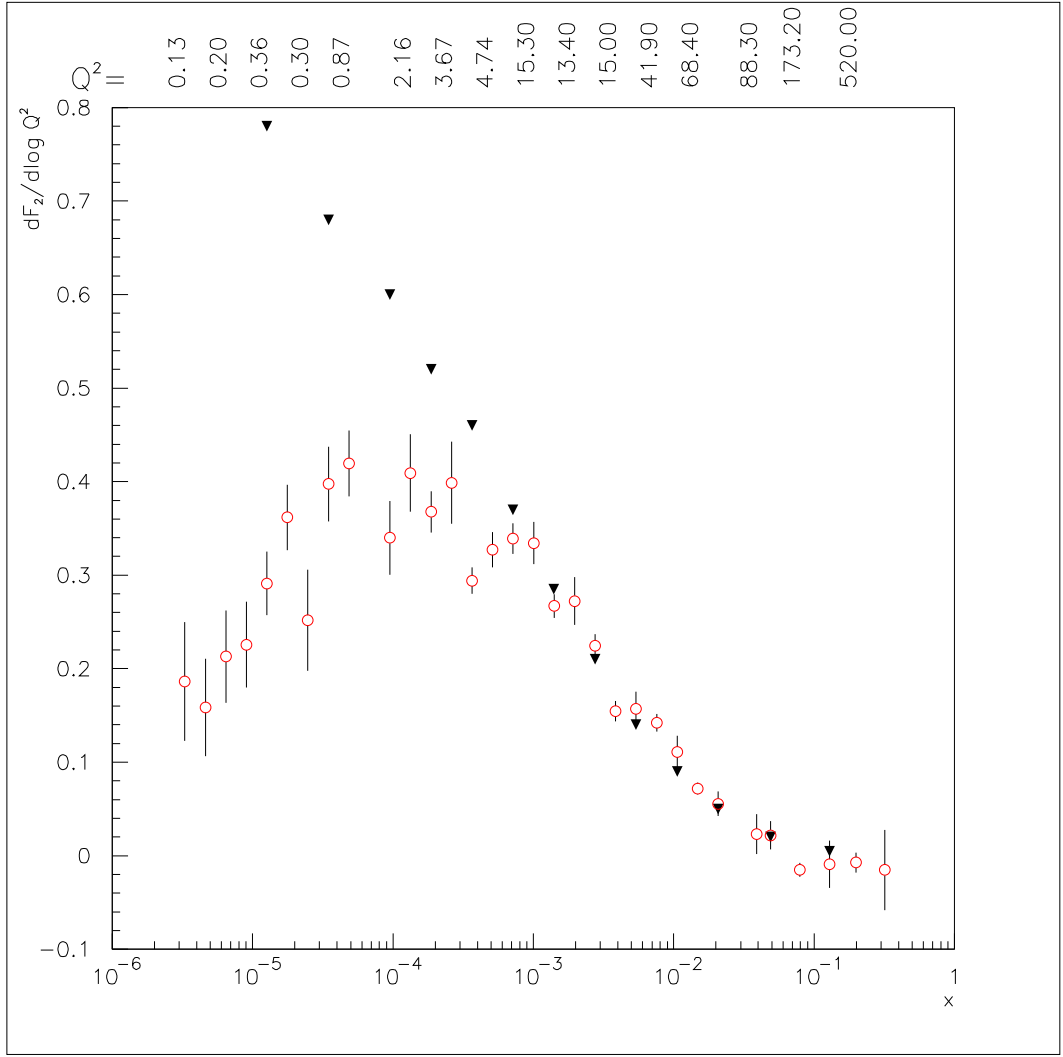,width=140mm}
\caption{The  HERA data and GRV'94 prediction (triangles) on the
$F_2(x,Q^2)$ slope. The data are taken from Ref.\cite{F2SLHERA}.}  
\label{Fig.1}
\end{figure}

 There are two different
types of SC 
that contribute to change of the $F_2$ slope:
\newline
(i) SC due to passage of the
quark - antiquark pair
through the nucleon, which lead to a more general equation for the $F_2$
slope than \eq{DGLAPSL} and   (ii) SC to the gluon
structure function
in
\eq{DGLAPSL}. We first discuss the quark-antiquark sector.

1.  Closed formulae for the penetration of a
 $q \bar q $ - pair through the target in the Eikonal ( Glauber )
approach were written many years ago \cite{LR87} \cite{MU90} and
have been discussed in detail  over  the past few  years \cite{QQSC}.
For the $F_2$ slope these formulae lead to
\beq \label{QQSC}
\frac{\partial F_2(x,Q^2)}{\partial \ln (Q^2/ Q^2_0)}\,\,=\,\,\frac{Q^2}{
3 \,\pi^2}\,\int
d b^2_t\,\,\{\,\,1\,\,\,-\,\,\,e^{- \kappa(x,Q^2; b^2_t)}\,\,\}\,\,,
\eeq
where 
\beq \label{KAPPA}
\kappa(x,Q^2; b^2_t)\,\,=\,\,\frac{2 \pi \as
}{3 Q^2}\,\Gamma(b_t)\,\,xG^{DGLAP}(x,Q^2 )\,\,.
\eeq
Here, we have used the main property of the DGLAP evolution equations,
which allows us to factor out the impact parameter (  $b_t$ ) dependence
from $x$ and
$ Q^2$  ( see Ref.\cite{GLMPH} ).  $Q^2_0$ is the photon   virtuality
scale from which we apply pQCD for our calculation.
 $\Gamma(b_t)$ denotes
the Fourier
transform of the two gluon nonperturbative  form factor of the target,
which is independent of the incident particle.
The relation between the  profile $\Gamma(b_t)$ and the two gluon form
factor
reads
\beq \label{BDEP}
\Gamma(b_t)\,\,=\,\,\frac{1}{\pi}\,\,\int\,e^{- i (\vec{b}_t
\cdot\vec{q}_t)}\,F(t)\,d^2 q_t
\eeq
with $t\,=\,-\,q^2_t$. Note, that factorization  in \eq{KAPPA} is valid
only for $|t|\,\leq\,Q^2_0$.
 To simplify the  calculations we
approximate 
\beq \label{B1}
\Gamma(b_t)\,\,=\,\,\frac{1}{R^2}\,e^{- \frac{b^2_t}{R^2}}\,\,.
\eeq

An  impressive property of \eq{QQSC} is the fact that the SC depend
on the gluon structure function at short transverse distances
$r^2_{\perp}\,=\,\frac{4}{Q^2}$. This  means that for the $F_2$ slope we
can
calculate the  SC with guaranteed theoretical accuracy in pQCD,  while the
SC
for the deep
inelastic structure function ( $F_2$  ) which  originate from large
distances
$r^2_{\perp}\,\geq \,\frac{4}{Q^2}$ and are, thus,
not well defined and could   lead to   errors in the
calculations.
\begin{figure}[htbp]
\epsfig{file=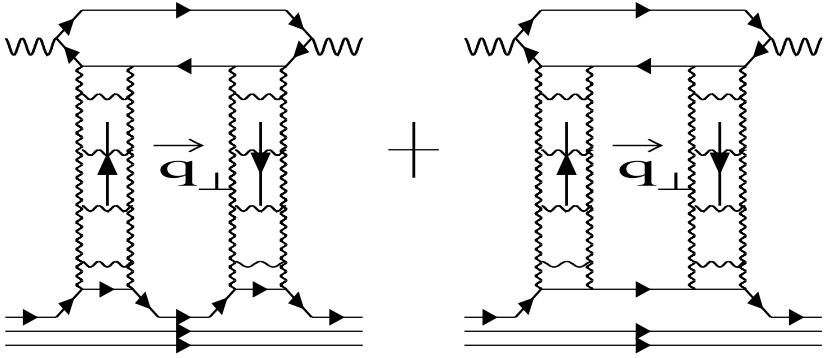,width=140mm,height= 65mm}
\caption{The first order SC $\propto \,\kappa^2$  for  $F_2(x,Q^2)$.}
\label{Fig.2}
\end{figure}

The nonperturbative QCD ( npQCD )  information we need is given  in 
 \eq{KAPPA} by  the nonperturbative  profile $\Gamma(b_t)$.  In
Fig.2 we 
show the lowest order  SC to $F_2$ in \eq{QQSC} which are proportional to
$\kappa^2$.
 In the additive quark model ( AQM ) we have two diagrams shown  in
Fig.2 which are of order $\kappa^2$.
 One can see that in AQM we have two scales for the  integration over
$q_{\perp}$: the distance between two constituent quarks in a nucleon
 ( the first diagram  in Fig.2 ) 
 and
 the size of the constituent quark ( the second diagram  in Fig.2)  ( see
Refs.\cite{TWORAD}\,\cite{GLMSLOPE}  for details 
).  \eq{QQSC} is the
simplest formula in which we can  assume that two gluons inside a nucleon
have
no other correlation than  their being   confined in a nucleon
with size $R$. A  more general formula for the two radii model of the 
nucleon was obtained in Ref. \cite{GLMSLOPE} . 
\newpage
For the $F_2$
slope 
$$
\frac{\partial F^{SC}_2(x,Q^2)}{\partial  \ln (Q^2/
Q^2_0)}\,=\,\frac{Q^2}{ 3\,\pi^2}\,\int
d b^2_t\,\,\{\,1\,\,-\,\,e^{- \kappa_1(x,Q^2; b^2_t)}\,\,+\,\,
\frac{\kappa^2_2(x,Q^2;b^2_t)}{\kappa_1(x,Q^2;b^2_t)\,-\,\kappa_2(x,Q^2;b^2_t)}
\,e^{-
  \kappa_1(x,Q^2;b^2_t)}
$$
\beq \label{QQTRSC}
+
\,\,\frac{\kappa^2_2(x,Q^2;b^2_t)}{(\kappa_1(x,Q^2;b^2_t)\,
-\,\kappa_2(x,Q^2;b^2_t))^2}\,(\,e^{-
\kappa_1(x,Q^2;b^2_t)}\,
\,-\,\,e^{- \kappa_2(x,Q^2;b^2_t)}\,)\,\,\}\,\,,
\eeq
where ( see Ref. \cite{GLMSLOPE} for details )
$$
\kappa_1(x,Q^2;b_t)\,\,=\,\,\frac{2 \pi \as}{3 Q^2 R^2_1}\,x
G^{DGLAP}(x,Q^2)
\,e^{- \frac{b^2_t}{R^2_1}}\,\,=\,\,\kappa_1(x,Q^2)\,e^{-
\frac{b^2_t}{R^2_1}};
$$
\beq \label{KAPPAS}
\kappa_2(x, Q^2; b_t)\,\,=\,\,\kappa_1(x,Q^2)\,\frac{R_1}{R_2}\,e^{ -
\frac{-b^2_t}{R^2_2}}\,\,.
\eeq
To evaluate the influence of the SC we introduce a damping factor (
$D(\kappa)$ )
\beq \label{DF}
D_Q( \kappa )\,\,=\,\,\frac{ \frac{\partial F^{SC}_2( x, Q^2)}{\partial
\ln(Q^2/Q^2_0)}}{\frac{Q^2}{ 3\,\pi^2}\,\int d b^2_t
\kappa(x,Q^2; b^2_t)}\,\,,
\eeq
where $\frac{\partial F^{SC}_2( x, Q^2)}{\partial
\ln(Q^2/Q^2_0)}$ is calculated from  \eq{QQSC} for a one radius model of
the nucleon  and from  \eq{QQTRSC} for a  two radii model. The denominator
is the first term of the $F_2$ slope which corresponds to the DGLAP
equations.
Note, that in such a calculation for a two radii model $R = R_1$. 
 
Using this damping factor $D (\kappa)$ we can write the $F_2$ slope in the
form
\beq \label{DFSL}
\frac{\partial F^{SC}_2( x,Q^2)}{\partial\ln(Q^2/Q^2_0)}\,\,=\,\,D_Q(
\kappa
)\,\,\frac{\partial F^{DGLAP}_2( x, Q^2)}{\partial
\ln(Q^2/Q^2_0)}\,\,.
\eeq

The calculated damping factor of \eq{DF} is plotted in Fig.3 for a  one
radius model with $R^2 = 10 GeV^{-2}$ \cite{GLMSLOPE} versus $\kappa$ (
upper curve ) and
for two a radii model with two sets of radii: (i) $R^2_1 = 10 GeV^{-2}$
and
$R^2_2 = 3 GeV^{-2}$ and (ii) $R^2_1 = 6 GeV^{-2}$ and $R^2_2 = 2
GeV^{-2}$  versus $\kappa_1$ ( see \eq{KAPPAS} ). We note that two radii
sets of curves are almost undistinguishable from one another as a function
of $\kappa_1$.

\begin{figure}[htbp]
\epsfig{file=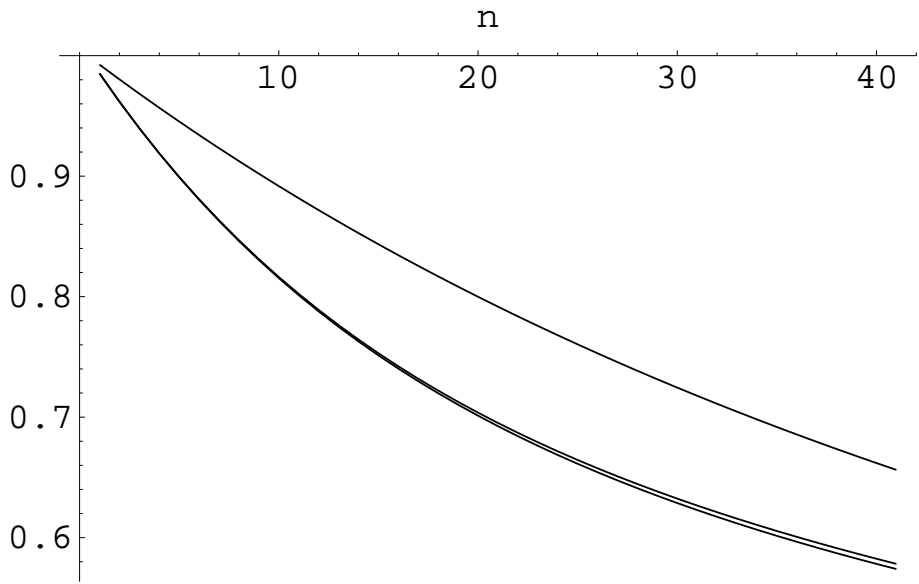,width=140mm}
\caption{The damping factor $ D_Q( \kappa )$ versus $\kappa (\kappa_1) =
0.03\, +\,
0.05\,n$ for a one radius model with  $R^2 = 10 GeV^{-2} $ ( upper curve)
and for a two radii model with two sets of radii: (i) $R^2_1 = 10
GeV^{-2}$
and $R^2_2 = 3 GeV^{-2}$ and (ii) $R^2_1 = 6 GeV^{-2}$ and $R^2_2 = 2
GeV^{-2}$.}
 
\label{Fig.3}
 \end{figure}

One can see that the two radii model of the nucleon leads to sufficiently
large  SC which depend on the set of radii chosen. Note that
the
value
of $\kappa_1$ is inversly  proportional to $R^2_1$.

2. The Glauber ( Eikonal
) formula for the SC in the gluon structure function  was obtained by
Mueller \cite{MU90} and
discussed in details in Ref.\cite{AGLA}.
\beq \label{GSC}
xG^{SC}(x,Q^2)\,\,=\,\,\frac{2}{
\pi^2}\,\int^1_x \,\frac{d x'}{x'} \,\int^{Q^2}_{Q^2_0} \,d Q'^2
\,\int
d b^2_t\,\,\{\,\,1\,\,\,-\,\,\,e^{- \kappa_G(x',Q'^2; b^2_t)}\,\,\}\,\,,
\eeq
with $\kappa_G( x',Q'^2;b^2_t)\,=\,\frac{9}{4} \kappa (x',Q'^2;b^2_t)$
where $\kappa$
is taken from \eq{KAPPA}.  In the case of a  two radii model the formula
can be derived as a direct generalization of the procedure suggested in
Ref. \cite{GLMSLOPE}
\beq \label{GTRSC}
 xG(x,Q^2) \,\,=\,\,\frac{2}{ \pi^2}\,\int^1_x\,\frac{d
x'}{x'}\,\int^{Q^2}_{Q^2_0}\,d Q'^2\,\int
d b^2_t\,\,\{\,\,1\,\,\,-\,\,\,e^{- \kappa_{G1}(x',Q'^2; b^2_t)}\,\,+
\eeq
$$
\frac{\kappa^2_{G2}(x',Q'^2;b^2_t)}{\kappa_{G1}(x',Q'^2;b^2_t)\,-
\,\kappa_{G2}(x',Q'^2;b^2_t)}\,e^{-
  \kappa_{G1}(x',Q'^2;b^2_t)}\,\,+
$$
$$
\,\,\frac{\kappa^2_{G2}(x',Q'^2;b^2_t)}{(\,\kappa_{G1}(x',Q'^2;b^2_t)
\,-\,\kappa_{G2}(x',Q'^2;b^2_t)\,)^2}\,(\,e^{-
\kappa_{G1}(x',Q'^2;b^2_t)}\,
\,-\,\,e^{- \kappa_{G2}(x',Q'^2;b^2_t)}\,)\,\,\}\,\,,
$$
where $\kappa_{G1}\,=\,\frac{9}{4} \,\kappa_1$ and
$\kappa_{G2}\,=\,\frac{9}{4}\,\kappa_2$ where $\kappa_1$ and $\kappa_2$
are defined in \eq{KAPPAS}.

 The
gluon damping factor is  defined  as
\beq \label{GDF}
D_G(x,Q^2)\,=\,\frac{xG^{SC}(x,Q^2)}{\frac{2}{\pi^2}\int^1_x\,\frac{d 
x'}{x'}\,\int^{Q^2}_{Q^2_0} \,d Q'^2 \,\int\,d b^2_t
\,\kappa_{G}(x',Q'^2;b^2_t)}\,\,,
\eeq
where $x G^{SC}$ is calculated from \eq{GSC} for a  one radius model and
from
\eq{GTRSC} for a two radii model. For a  two radii model $\kappa_G =
\kappa_{G1}$
in the dominator of \eq{GDF}.
\alphfig
\begin{figure}[htbp]
\centerline{\epsfig{file=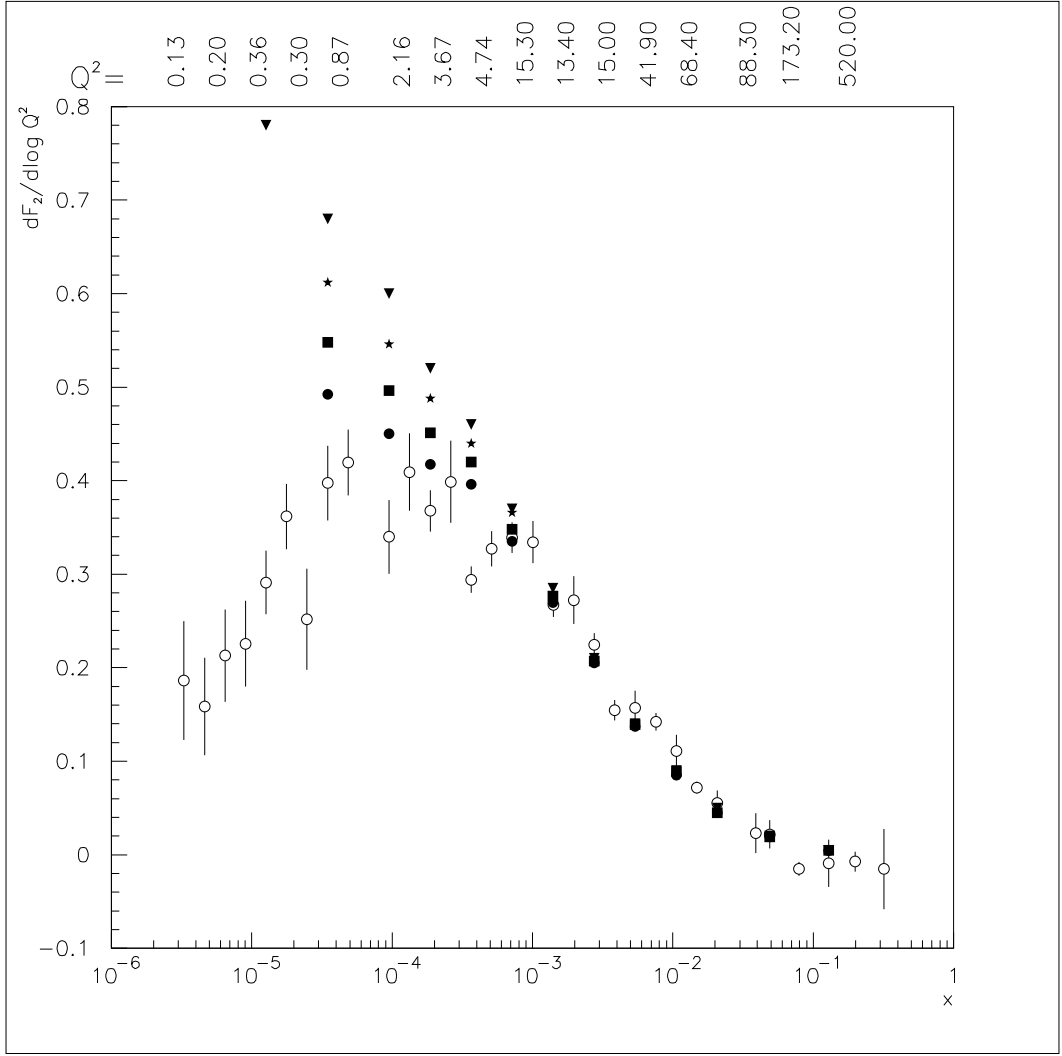,width=140mm}}
\caption{The $F_2$ slope for SC in quark sector only. Triangles are the
GRV'94 prediction. Stars are  the result of SC calculations in the one
radius model for a nucleon with $R^2 = 10 GeV^2$, squares are  for the
SC in the two
radii model with  $R^2_1 = 10 GeV^2 $ and $R^2_2 =
3GeV^2$ and  circles are for the SC in  the  two radii model with $R^2_1 =
6GeV^2$ and
$R^2_2 = 2 GeV^2$. }
\label{fig4a}
 \end{figure}
It is important to note that the $Q'^2$ integration in \eq{GTRSC} and
\eq{GDF} spans all distances, including  long distances dominated by
nonperturbative dynamics. This is very different from \eq{QQSC} and
\eq{QQTRSC} where the SC depend on the gluon density at short transverse
distances. Since a theoretical approach to npQCD 
is still lacking, we eliminate the long distance contributions to \eq{GSC}
and \eq{GTRSC} by imposing a low cutoff on the $Q'^2$ integration. With
this cutoff we neglect the contributions due to transverse distances
$r_{\perp}\,>\,\frac{1}{Q_0}$. This procedure makes our calculation of
$D_G(x,Q^2)$ less reliable than our calculation of $D_Q(x,Q^2)$. To
evaluate   the errors due to this source we have calculated $xG(x,Q^2) $
with two different cutoff values: $Q^2_0 = 0.4 \,GeV^2$ and $Q^2_0 = 1
\,GeV^2$, using the GRV'94 parameterization \cite{GRV} for the  solution
of the DGLAP evolution equations. The resulting gluon damping factors,
$D_G(x,Q^2)$ differ by about 10\% which lends a reasonable credibility to
our calculation.  As we have noted in the introduction, we are of the
opinion that the GRV'94 parameterization does not contain   SC. This is
due to the fact that most of the experimental sample used to fix the
GRV'94 parameters has values of $x\,>\,10^{-3}$, where we estimate the SC
are very small.

\begin{figure}[htbp]
\centerline{\epsfig{file=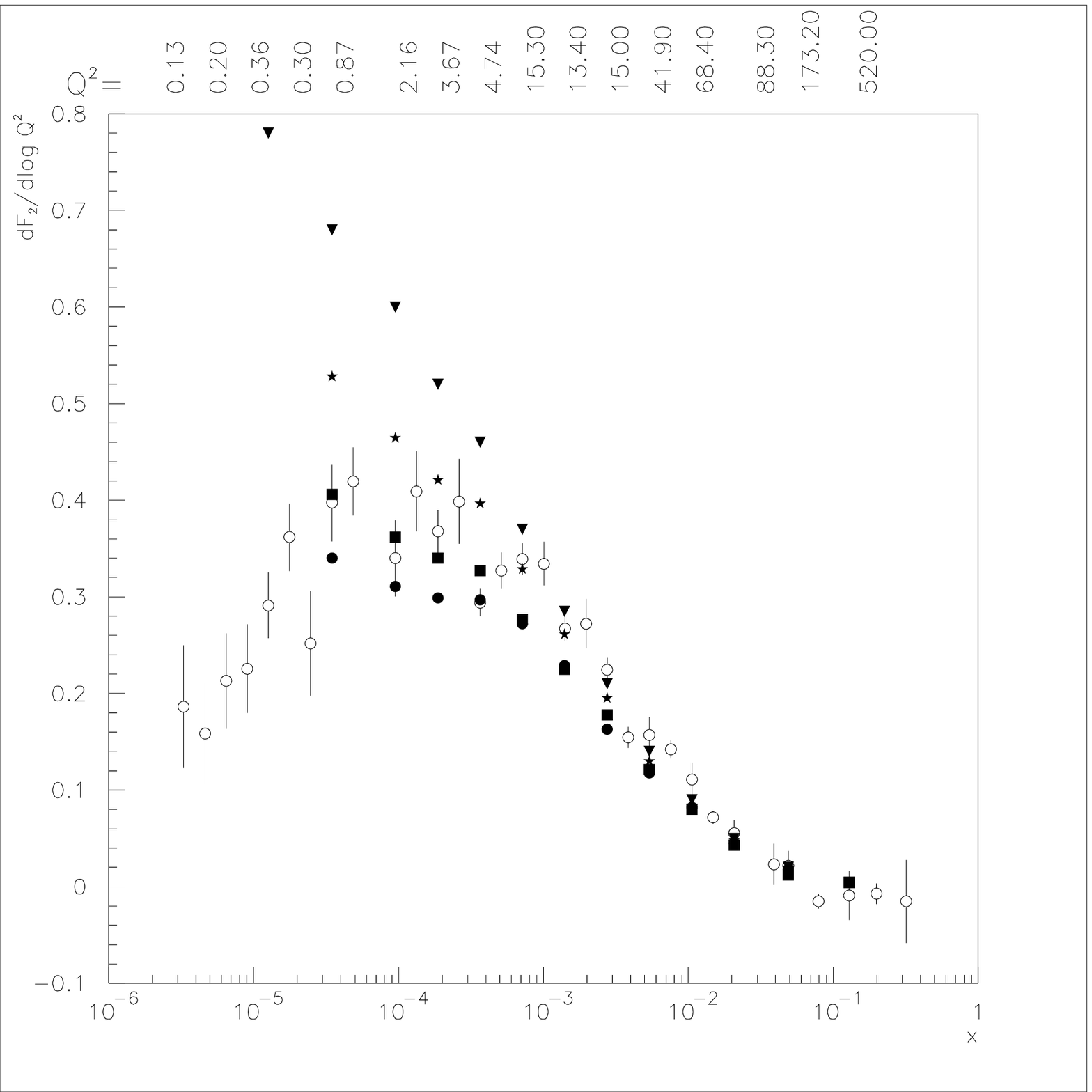,width=140mm}}
\caption{The $F_2$ slope for SC both in quark and gluon sectors.
Notations are the same as in Fig.4-a. }
\label{fig4b}
 \end{figure}
\resetfig

The formula which we use to  compare  with HERA experimental data is
\beq \label{FIN}
\frac{\partial F^{SC}_2(x,Q^2)}{\partial\ln(Q^2/Q^2_0)}
\,\,=\,\,D_Q(x,Q^2)\,\,D_G(x,Q^2)\,\frac{\partial
F^{GRV}_2(x,Q^2)}{\partial\ln(Q^2/Q^2_0)}\,\,,
\eeq
where $F^{GRV}_2$ is the deep inelastic structure function calculated in
the DGLAP evolution approach with  the GRV'94  parameterization.
Our results  compared with the experimental data
\cite{HERALOWQ} are shown in Fig.4 . Since we use the GRV'94 input our
calculations can be
carried out only for $Q^2\,>\, 0.4\, GeV^2$.

To summarize, the main points of this letter are:

1) The deviation of $ \frac{\partial 
F^{SC}_2(x,Q^2)}{\partial\ln(Q^2/Q^2_0)}$ from the behaviour predicted by
the DGLAP evolution in the small $Q^2$ and exceedingly small $x$ region is
associated with SC applied both to the passage of  quark - antiquark pair
through the nucleon and the gluon density within the nucleon target.
When SC are applied good agreement with the new HERA data, with $Q^2
\,>\,0.8\,GeV^2 $ , is obtained.

2) SC for the quark - antiquark interaction with the target nucleon is
concentrated at short distances $r_{\perp} \,\approx\,\frac{2}{Q}$
and, therefore, can be calculated to a good degree of accuracy in pQCD.
The effects of these SC are sufficiently large to account for most of the 
difference between the DGLAP prediction and experimental data,,  unlike
the
case of the deep inelastic structure function $F_2$ \cite{TWORAD}.

3) The calculated SC for the gluon structure function are large and
contain  uncertainties due to long distance contributions which have not
been included in the calculation. At present it is not possible  to
evaluate the errors which arise   from   nonperturbative contribution. We
have checked  the relative contribution coming from
$3\,GeV^2\,\geq\,r^2_{\perp}\,\geq\,1\,GeV^2$  by changing $Q^2_0$ - the
lower limit of $Q'^2$ integration in the calculation of $xG^{SC}(x,Q^2)$.
The resulting change in $D_G(x,Q^2)$ is not more than 10\%.

4)  Our determination of the two radii of  the nucleon rests on
J/$\Psi$ photo and DIS production data \cite{GLMSLOPE}. The present
analysis
suggests that better data on the $F_2$ logarithmic $Q^2$ slope may
provide an independent determination of these radii as well as additional
knowledge on the role of long distance nonperturbative contributions to
the SC.  

{\bf Acknowlegements:} We wish to thank A. Caldwell and A. Levy for
providing us with both the data  and GRV'94 points of Figs. 1 and 4. 
This research was supported in part by the Israel Science Foundation
founded by the Israel Academy of Science and Humanities.

\end{document}